\documentclass[12pt]{article}
\usepackage{epsfig}
\input{psfig}

\newlength{\horilen}

\newcommand{\figcap}[2]
{\begin{center}
\raisebox{-0.8mm}{\parbox[b]{2.5cm}
{\caption{}\label{#1}}}
\parbox[t]{12cm}{{\sl #2}}
\end{center}}

\def\openone{\leavevmode\hbox{\small 1\kern-3.8pt\normalsize 1}}

\def\Cfat{\mbox{\sl C\kern-12.0pt C}}
\def\Ccmplx{\leavevmode\hbox{$C$\kern-1.4pt $\ell_{0,1}$}}
\def\Cquate{\leavevmode\hbox{$C$\kern-1.4pt $\ell_{0,2}$}}
\def\Cpauli{\leavevmode\hbox{$C$\kern-1.4pt $\ell_{3,0}$}}
\def\Cmajor{\leavevmode\hbox{$C$\kern-1.4pt $\ell_{3,1}$}}
\def\Csptim{\leavevmode\hbox{$C$\kern-1.4pt $\ell_{1,3}$}}
\def\Cdirac{\leavevmode\hbox{$C$\kern-1.4pt $\ell_{4,1}$}}

\newcommand{\yhat}{\kern1.2pt\hat{\mbox{\kern-1.2pt\boldmath $y$}}}

\newcommand{\jhat}{\kern2.0pt\hat{\mbox{\kern-2.0pt\boldmath $\jmath$}}}

\newcommand{\eqbeg}{\protect \begin{equation}}
\newcommand{\eqend}{\protect \end{equation}}

\newcommand{\head}[1]{\large\bf~\\{#1}\normalsize\rm \vspace{2mm}}

\begin{document}
\setcounter{page}{1}
\setcounter{equation}{0}
\setcounter{figure}{0}

\hspace*{6.5cm} {\sl Published in: Van A tot Q, NNV, November, 2000}\\ 
\begin{center}
~\LARGE\bf
Light is Heavy\\
~\large\rm\\
M.B. van der Mark and G.W. 't Hooft \\
~\normalsize\sl\\
Philips Research Laboratories,\\
Prof.\ Holstlaan 4, 5656 AA~ Eindhoven, The Netherlands\\~\\
\rm\end{center}

\begin{abstract}
\noindent
Einstein's relativity theory appears to be very accurate, but at times equally 
puzzling. On the one hand, electromagnetic radiation must have zero rest mass in 
order to propagate at the speed of light, but on the other hand, since it 
definitely carries momentum and energy, it has non-zero inertial mass. Hence, by 
the principle of equivalence, it must have non-zero gravitational mass, and so, 
light must be heavy. In this paper, no new results will be derived, but a 
possibly surprising perspective on the above paradox is given.
\end{abstract}

\head{Introduction}
~\\
Einstein's general theory of relativity is based on two experimental facts. 
First, that the speed of light appears to be equal for all observers, 
independent of their velocity with respect to the light source, and second, the 
apparent equality of gravitational mass $m_g$ and inertial mass $m_i$ 
\cite{sch99}. The latter is expressed by what is known as the principle of 
equivalence: ``No experiment can distinguish the effects of a gravitational 
force from that of an inertial force in an accelerated frame". (Actually, only 
proportionality instead of equality between $m_g$ and $m_i$ is found, but the 
proportionality constant can be taken unity).

\head{What are inertial mass and gravitational mass?}
~\\
The inertial mass $m_i$ is a measure of persistence to stay in the same state of 
motion, or like a resistance to acceleration, expressed by Newton's law 
$F=m_ia$. The inertial mass of an object can be determined, for example, by 
measuring the changes of velocities in a collision with another object of known 
mass. 
Gravitational mass $m_g$ is a measure for attraction of and attraction by other 
masses $M_g$, it is like a ``gravitational charge'', the force being 
$F=GM_gm_g/r^2$. The gravitational mass of an object can be determined by 
putting it on a balance to compare it with a reference mass.

\head{The paradox: light and heavy light}
~\\
Electromagnetic radiation carries momentum $\vec{p}=\hbar\vec{k}$ and energy 
$E=\hbar\omega=pc$, and it can exert a (radiation) pressure on any object it 
falls upon. Hence, light has inertia that can be quantified by an inertial mass 
$m_i=\hbar k/c$. Now, the principle of equivalence tells us that light must also 
have a gravitational mass $m_g$, and consequently it must be attracted by heavy 
bodies. That this is the case is, of course, well known from the bending of 
starlight as observed during solar eclipse experiments as well as from the 
gravitational Doppler shift of light as seen in a vertical gamma-ray 
spectrometer employing the M\"ossbauer effect. If we would like to verify for 
electromagnetic radiation that indeed $m_g=m_i$, the question arises as how to 
determine the magnitude of its gravitational mass. For an arbitrary object, one 
would normally weigh the object, simply by putting it on a scale, at rest, thus 
measuring its rest mass $m_0$. From this, it would appear that the rest mass and 
gravitational mass are the same thing. But how does one weigh light? It usually 
flies off with $c$, the speed of light!  If, nevertheless, we would be able to 
accomplish this speedy task, we would find the mass of light to be zero. This 
can be seen as follows. Consider the Lorentz transformation of the inertial mass  
$m_i=\gamma m_0$ of an object with rest mass $m_0$ and velocity $v$ (where 
$\gamma=1/\sqrt{1-v^2/c^2}$). In the limiting case of light-speed velocity, 
$v=c$, $m_i$ becomes infinite unless $m_0=0$. Since light has finite inertia, 
the rest mass of light should be zero. This then seems to be in contradiction: 
on the one hand $0\neq m_i=m_g$ and on the other $m_g=m_0=0$. The questions that 
emerge are: How to weigh anything properly? and: What is rest mass?

\head{How to weigh a gas}
~\\
How to properly weigh something as volatile as a gas? Simply put it in a box so 
that it doesn't fly away, and then put it on a symmetric balance. The reference 
mass should be put in a box of the same size to eliminate differences in up lift 
by the surrounding atmosphere.
To make this plausible, consider, for example, a gas at temperature $T$ and 
pressure $P$ of $N$ particles of mass $m=\gamma m_0$ inside a rectangular box of 
height $h$ and volume $V=Ah$, with $A$ the area of both its top and bottom. Due 
to gravitation, the pressure of the gas decays exponentially with height:
\eqbeg
\label{Pressure}
P(h)=P(h=0)e^{-h\frac{mg}{kT}} 
\eqend
More gas particles collide faster with the bottom than with the top of the box, 
resulting in a pressure difference between the top and bottom ($h=0$), given by 
$P(h)-P(h=0) \approx -P(h=0)hmg/kT$. Strictly speaking, this is only true if $h$ 
is much smaller than the characteristic height $h_c=kT/mg$ which is, for 
example, $8.5\:$km for the atmosphere. For an ideal gas $P=kTN/V$ and the force 
is $F=PA$, hence the net force on the box is
\eqbeg
\label{Force}
F=-Nmg 
\eqend
Indeed, $M=Nm$ is what we would expect for both the inertial and the 
gravitational mass of the gas. Note that the mean velocity of the particles, and 
hence $\gamma$, increases with temperature. So, on weighing the box, we exactly 
do find the relativistic mass $M=Nm=N\gamma m_0$ of the gas, i.e.\ the rest mass 
of the particles plus the mass represented by their kinetic energy. The hotter 
the gas, the heavier the contents of the box. Fluctuations in pressure at time 
scales of the inverse collision frequency will be dampened/averaged out by the 
inertia of the balance such that a stable reading of the mass, the gravitational 
mass of the closed system, is obtained. Although nothing at all is at rest 
inside the box, the gravitational mass is equal to the rest mass of the box as a 
whole! The rest mass of the box is not equal to the sum of the rest masses of 
its contents. The reason that we find $M_g=M_0$ is that we have a closed system 
with the {\sl centre of mass} of all the particles at rest.

\head{Light on the balance}
~\\
The same box, but now filled with light, and with the inner walls made perfectly 
reflecting, can be weighed too. Similar to the case where it was filled with gas 
particles, the light or, if you like it better, the photons, are gravitationally 
red or blue shifted at upward or downward propagation respectively. This again 
results in a net (radiation) pressure on the balance \cite{fey89}. The shorter 
the wavelength, $\lambda=2\pi\hbar c/E$, of the photons, the heavier the box. 
From the outside, it is {\sl impossible} to judge whether the box is filled with 
a simple gas or with light.
\begin{figure}[h] 
\setlength{\horilen}{13.0cm} 
\vspace{-5mm}
\begin{center}
\begin{tabular}{c}
\psfig{figure=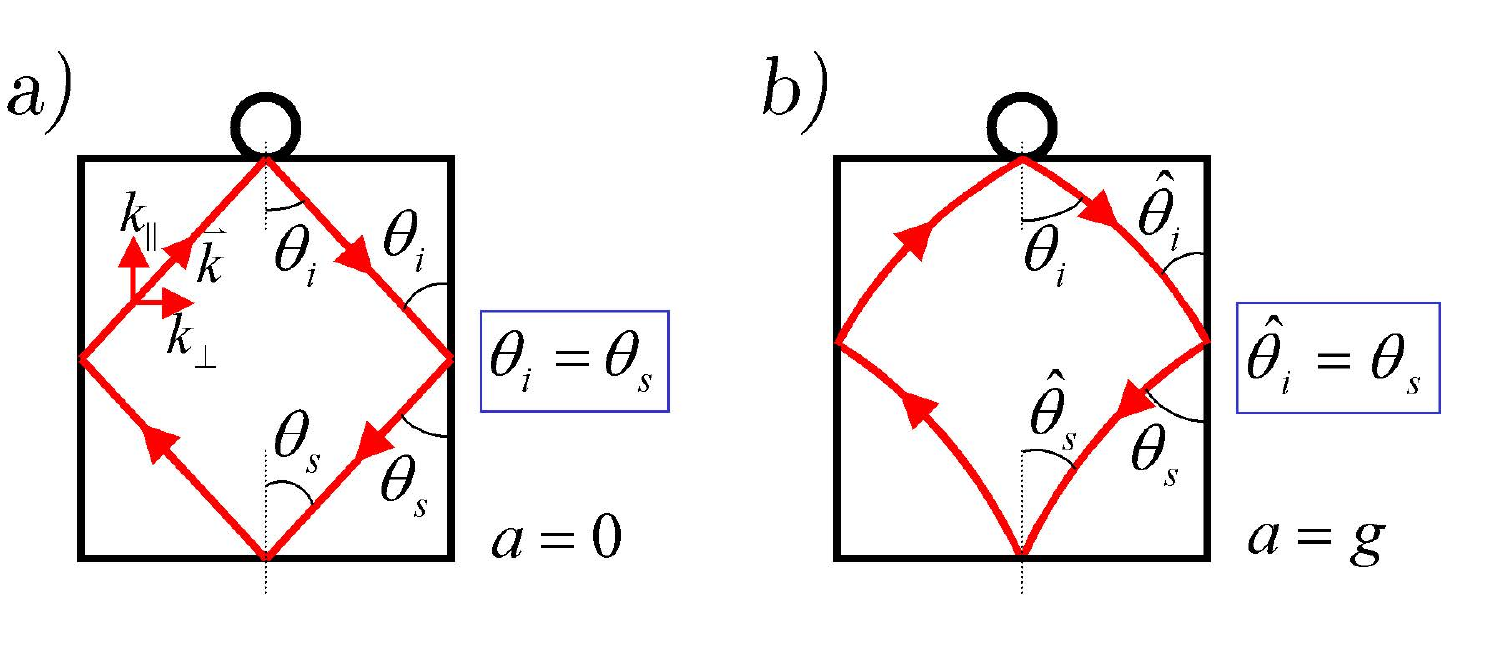,width=\the\horilen}
\end{tabular}
\end{center}
\vspace{-15mm}
\figcap{boxes1}{Light in a reflecting box, a) in free fall, b) in a strong
gravitational field.}
\end{figure} 
\begin{figure}[ht] 
\setlength{\horilen}{13.0cm} 
\vspace{-5mm}
\begin{center}
\begin{tabular}{c}
\psfig{figure=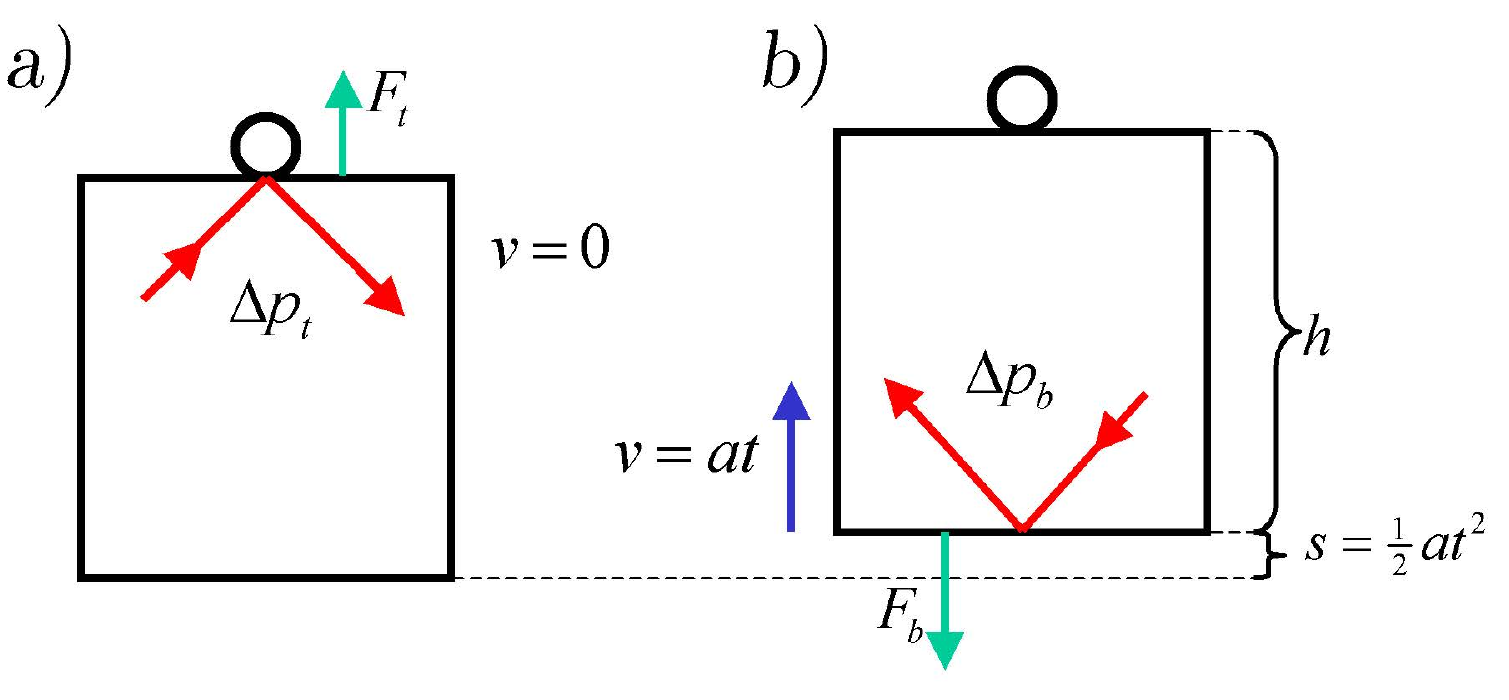,width=\the\horilen}
\end{tabular}
\end{center}
\vspace{-15mm}
\figcap{boxes2}{Snapshots of Fig.~\ref{boxes1}b, the momentum transfer at
top and bottom, after application of the equivalence principle, both seen
from the same frame.} 
\end{figure} 
For a proof, compare the two drawings in Fig.~\ref{boxes1}, both showing light 
circulating with round trip time $\Delta t=t_{up}+t_{down}=2t$ inside a 
reflecting box, where $t$ is the time required to travel from top to bottom, or, 
because the speed of light is constant, vice versa. The first box is floating 
freely in space, the second is at rest on a heavy planet, or alternatively, is 
accelerated in deep space with $a=g$. In both cases, the observer and box are in 
the same frame. The wave vector $\vec{k}$ and frequency $\omega$ of the light 
are related as follows: $k=\omega/c$, $\cos\theta=k_{\parallel}/k$ and
\eqbeg
\label{wavevect}
k^2_{\parallel} + k^2_{\perp}-\frac{\omega^2}{c^2} =0 
\eqend
To calculate the net radiation force on the box, we will consider the reflection 
of a ``photo'' from top and bottom of the box separately, see snapshots of 
Fig.~\ref{boxes1}b in Fig.~\ref{boxes2}. The mass of the box is defined in the 
rest frame, $v=0$, which we define with respect to be its point of suspension, 
the top, where we let the ``photon'' strike first. Now, we employ the 
equivalence principle on the bottom: we consider the gravitational force to come 
from an acceleration of the box upwards. The result is that the bottom seems to 
have velocity $v=at$ with respect to the top at the moment $t_{down}=t$ of 
impact of the ``photon''. The Doppler shift of the light can be calculated using 
the following Lorentz transformations for $k$ and $\omega$ in the rest frame to  
$\hat k$ and $\hat\omega$ in a frame moving with velocity $v$:
\begin{eqnarray}
\label{Doppler}
\hat k_{\parallel} &=& \gamma\left( k_{\parallel}-\frac{v}{c^2}\omega \right) \\
\hat k_{\perp} &=& k_{\perp} \\
\hat \omega &=& \gamma\left( \omega - vk_{\parallel} \right)
\end{eqnarray}
From the above, it follows that we only need to consider those components of $k$  
which are (anti-)parallel to the acceleration. The total momentum transfer of 
light on top and bottom results in a net force:
\eqbeg
\label{impulse}
\vec{F} = \vec{F_b} + \vec{F_t}= \frac{\Delta \vec{p_b} + \Delta 
\vec{p_t}}{\Delta t} 
\eqend
The momentum transfer during one roundtrip is calculated in the momentary frame 
of the box (for downward flight $k_{\parallel}\longrightarrow -k_{\parallel}$):
\begin{eqnarray}
\label{momtrans}
\Delta p_t &=& 2\hbar k_{\parallel} \\
\Delta p_b &=& \frac{2}{\gamma}\hat p_b = \frac{2}{\gamma} \hbar\hat 
k_{\parallel} = 2\hbar\left( -k_{\parallel}-\frac{v}{c^2}\omega \right)
\end{eqnarray}
Substituting those in Eq.~(\ref{impulse}), and using that $\Delta t=2t$, $v=at$, 
$E=\hbar\omega$ and $a=g$ we find that
\eqbeg
\label{totalforce}
F = -\frac{\hbar\omega}{c^2} \frac{2v}{\Delta t} 
= -\frac{E}{c^2}a = -m_g g 
\eqend
which concludes the proof that the gravitational mass of light is $m_g=E/c^2$. 
See also Ref.~\cite{fey89}.

\head{Guess who?}
~\\
A combination of both the gas and light examples presented above is offered by 
the dramatic event of electron-positron pair annihilation. In the simplest case, 
just two photons are produced. Matter is fully transformed to radiation, but the 
mass stays. Put on a balance in a box, it is impossible to know whether or not 
the pair has decayed.
This example shows that the equation $E=mc^2$ expresses the equivalence of mass 
and energy and {\sl not} the generation of energy as a reaction product from 
mass. The confusion that sometimes arises can often be traced back to the mix-up 
between the words ``mass'' and ``matter''. Matter can be transformed into 
radiation. Matter is taking the role of energy container, radiation is some sort 
of released, ``free'' energy, that must fly through space.

\head{Discussion}
~\\
In the case of light, the rest mass is zero, but the gravitational mass equals 
the inertial mass, which is identical to the relativistic mass. The ``photon'' 
can only be weighed if it is contained in one way or another, so that its centre 
of mass is fixed (on average). 
In case we weigh any material object, heat, rotational, vibrational and kinetic 
energy, the sort of energy naturally contained in matter, put their weight to 
the scale. It shows that the term ``rest mass'' really only means that the 
centre of mass of the object is at rest in the frame of the observer. 
We can think of material objects as being built out of some smaller 
constituents, glued together by some binding force. We go from houses to bricks, 
from bricks to molecules, from molecules to atoms, from atoms to nucleons and 
electrons, and from there to quarks and still electrons (we could have started 
from cosmic super clusters).
From this list it should dawn on us that, every time we think, at first glance, 
that we are dealing with a rigid chunk of matter (planet, brick, atom), it 
appears to carry a lot of dynamics at various length scales and energies. The 
smaller the length scales, the stronger the forces involved and the higher the 
(binding) energies, and hence the corresponding masses, relative to the rest 
masses of the constituents. We could wonder whether this finds it climax at a 
point where an elementary material particle is build of constituents that have 
zero rest mass, with only kinetic and potential energy to make up for its mass. 
That this should be the case for the electron, but at the same time seems quite 
impossible \cite{fey64}, is well known \cite{wil00}.
What is intriguing is that matter's most basic building blocks, the elementary 
particles, all have non-zero spin, intrinsic angular momentum, which seems to 
imply that they all must have some sort of intrinsic dynamics. Hypothetical 
structures which do not have internal dynamics, such as point particles and hard 
spheres, do not exist. So what is matter really made of then? In the Dirac 
theory, the electron is like electromagnetic energy quivering at light speed, 
just like a photon in a box \cite{wil97}. If really so, matter is light.

\head{Conclusions}
\vspace{-2mm}
\begin{itemize}
\item Rest mass never applies to a system at complete rest, because such systems 
do not exist; there will {\sl always} be internal dynamics.
\item Rest mass applies to the centre of mass of a closed system
\item The gravitational mass is equivalent to the total energy of an object or 
system.
\item The mass of a closed system is always conserved. This is just the energy 
conservation law rephrased. Mass and energy are equivalent.
\end{itemize}
One could say: ``Matter is just ``canned'' energy, a box with internal dynamics, 
and radiation is ``free'' energy.''
If the photon would be put to rest, its gravitational mass would equal its rest 
mass, and hence vanish. The intriguing question is, what would happen if we 
could stop the electron from spinning?



\end{document}